\documentclass[12pt]{article}
\usepackage{times,amsmath,amsfonts,amsthm,amssymb,eucal}
\usepackage{algorithm}
\usepackage{algorithmic}
\usepackage{graphicx,ifthen}
\usepackage{wrapfig}
\usepackage{latexsym}
\usepackage{color}
\usepackage{mathrsfs}
\usepackage{bm}
\usepackage{bbm}
\DeclareGraphicsExtensions{.gif,.pdf,.png,.jpg,.tiff}
\begin{document}
\renewcommand{\baselinestretch}{1.38}
\def\Quote{\begin{quotation}\normalfont\small}
\def\EndQuote{\end{quotation}\rm}
\def\BigHeading{\bfseries\Large}\def\MediumHeading{\bfseries\large}
\def\bct{\begin{center}}
\def\ect{\end{center}}
\font\BigCaps=cmcsc9 scaled \magstep 1
\font\BigSlant=cmsl10    scaled \magstep 1
\def\lbk{\linebreak}
\def\Report{McJack for SAE after MS}
\def\Author{Jiang, Lahiri and Nguyen}
\pagestyle{myheadings}
\markboth{\Author}{\Report}
\thispagestyle{empty}
\bct{\BigHeading A Unified Monte-Carlo Jackknife for Small Area
Estimation after Model Selection}\\\vskip10pt
\BigCaps Jiming Jiang\lbk
\BigSlant University of California, Davis, U.S.A., jimjiang@ucdavis.edu\lbk
\BigCaps P. Lahiri\lbk
\BigSlant University of Maryland, College Park, U.S.A.\lbk
\BigCaps Thuan Nguyen\lbk
\BigSlant Oregon Health \& Science University, Portland, U.S.A.
\ect
\Quote
\vskip-5pt
We consider estimation of measure of uncertainty in small area
estimation (SAE) when a procedure of model selection is involved
prior to the estimation. A unified Monte-Carlo jackknife method,
called McJack, is proposed for estimating the logarithm of the
mean squared prediction error. We prove the second-order
unbiasedness of McJack, and demonstrate the performance of McJack
in assessing uncertainty in SAE after model selection through
empirical investigations that include simulation studies and
real-data analyses.
\vskip5pt\noindent\sl Key Words. \rm Computer intensive, Jackknife,
log-MSPE, measure of uncertainty, model selection, Monte-Carlo,
second-order unbiasedness, small area estimation.
\EndQuote
\section{Introduction}
\hspace{4mm}
Small area estimation (SAE) has become a very active area of statistical
research and applications. Here the term small area typically refers to
a population for which reliable statistics of interest cannot be
produced based on direct sampling from the population due to certain
limitations of the available data. Examples of small areas include a
geographical region (e.g., a state, county, municipality, etc.), a
demographic group (e.g., a specific age $\times$ sex $\times$ race
group), a demographic group within a geographic region, etc. See, for
example, Rao and Molina (2015) for an updated, comprehensive account
of various methods used in SAE. Statistical models, especially mixed
effects models, have played key roles in improving small area estimates
by borrowing strength from relevant sources. Therefore, it is not
surprising that model selection in SAE has received considerable
attention in recent literature. See, for example, Jiang, Nguyen and
Rao (2010), Datta, Hall and Mandal (2011), Pfeffermann (2013), Lahiri
and Suntornchost (2014), and Rao and Molina (2015).

The errors from model selection are likely to affect the uncertainty
measures in SAE estimates. To elaborate this point, let us consider
a specific aspect of model selection---inclusion of small area
specific random effects. Should one include area specific random
effect in small area modeling? Such a component is a compromise
between area specific fixed effects and no area effect and helps
improving the properties of model-based estimators. For example,
without such an area specific random effect, model-based estimator
may not be design-consistent, which may result in model-based estimate
for an area with large sample size to deviate significantly from the
corresponding design-based estimate, especially if area specific
auxiliary variables fail to capture variation across the areas. A
decision to exclude small area specific random effect may be based on
a significance test. But such a decision is anything but perfect and
depends very much on the subjective choice of the prespecified level
of significance. A reasonable uncertainty measure estimator must
incorporate the impact of model selection. However, most of the
uncertainty measure estimators, with the exception of Molina, Rao
and Datta (2015), do not attempt to capture the variation due to the
model choice and there is no analytical study to examine the important
second-order unbiasedness property of any of these estimators,
including that of Molina et al. (2015). In this paper, we propose a
new uncertainty measure of any small area model-based estimator that
incorporates errors due to model selection and a Monte-Carlo jackknife
second-order unbiased estimator of the proposed uncertainty measure.
We propose to use the logarithm of the mean squared prediction error
(MSPE) as the uncertainty measure, where MSPE incorporates errors
due to model selection. Our rationale behind using the log-MSPE comes
from the way lack-of-fit measure of a typical model selection criterion
is constructed. To elaborate on this point, consider the case of
regression model selection with normal data. The well-known
information criteria take the form of
\begin{eqnarray}
n\log(\hat{\sigma}^{2})+\lambda_{n}|M|,
\end{eqnarray}
where $n$ is the sample size, $\hat{\sigma}^{2}$ is the standard
estimator of the error variance, $\sigma^{2}$, $|M|$ is the dimension
of the model, $M$, typically defined as the number of free parameters
under $M$, and $\lambda_{n}$ is a penalty function. Thus, in this case,
the measure of lack-of-fit is proportional (under a fixed sample size)
to the logarithm of a variance estimator. Note that, typically, the
variance is of the same scale as the MSPE. Therefore, it is reasonable
to consider the logarithm of the MSPE as a measure of uncertainty in
SAE when a model selection procedure, such as an information criterion,
is involved. 

Besides the intuitive link to model selection, there are other
advantages of using the log-MSPE as a measure of uncertainty. In the
SAE literature, MSPE estimates have been routinely used in assessing
an improvement of the empirical best linear unbiased predictor (EBLUP)
over the direct estimator. For such a purpose, one can equivalently use
the log-MSPE, and report the improvement in the log-scale. An advantage
of log-MSPE over MSPE occurs when it is desirable to model uncertainty
measure estimators. This is because one can reasonably assume normality
of the error term when log-MSPE estimators are considered. Zimmerman
{\it et al.} (1999) emphasized the need to model log-MSPE in the context
of a geo-spatial application. Gershunskaya and Dorfman (2013) considered
modeling of logarithm of variances in an application related to Current
Employment Statistics survey. In a small area context, such a model can
provide a guideline for making important decisions on the choice of
different design factors (e.g., sample size, number of clusters) for a
future survey in achieving, approximately, a certain desired level of
log-MSPE of the proposed predictor for different small areas. Also,
the model can be used for quickly producing uncertainty measures when
it is time consuming to compute such measures when dealing with big
data as well as computational comlexity to meet a tight production
deadline.

In terms of statistical inference, it is easier to carry out
hypothesis testing when considering log-MSPE. For example, suppose
that one wishes to compare ${\rm MSPE}_{1}$ with ${\rm MSPE}_{2}$,
which may correspond to two different methods of SAE. If one has
second-order unbiased estimators of the log-MSPEs, say, $\hat
{l}_{j}$ for $l_{j}=\log({\rm MSPE}_{j})$, $j=1,2$, it is possible
to construct a z-test, or t-test, by assuming (approximately) that
$\hat{l}_{j}=l_{j}+e_{j}, j=1,2$, where $e_{j}$ is normal with mean
zero and constant variance.
 
Finally, a desirable property for an MSPE estimator is that it
needs to be positive. If the property is combined with the
second-order unbiasedness property, it turns out that it is very
difficult to produce an estimator that has both of these properties.
Typically, it is relatively easy to obtain a positive MSPE
estimator that is first-order unbiased. To achieve the
second-order unbiasedness, either analytical (e.g., Prasad and
Rao 1990) or resampling (e.g., Jiang, Lahiri and Wan 2002, Hall
and Maiti 2006) methods are used. However, with very few exceptions
(Prasad and Rao 1990, Chen and Lahiri 2011), these techniques do
not produce MSPE estimators that are guaranteed positive, in spite
of achieving the second-order unbiasedness. To ensure that the
MSPE estimator is positive, some modification of the (second-order
unbiased) MSPE estimator is often made. For example, Hall and Maiti
(2006) suggested the following strategy. Let $\widehat{\rm MSPE}_{1}$
and $\widehat{\rm MSPE}_{2}$ be two estimators of the same MSPE, for
example, the former being an MSPE estimator with a additive
bias-correction, and the latter one with a multiplicative
bias-correction. Both MSPE estimators have some types of problems.
For example, $\widehat{\rm MSPE}_{1}$ can take negative values, and
$\widehat{\rm MSPE}_{2}$ can be unreliable (Hall and Maiti 2006).
The idea is to combine the two estimators by letting $\widehat{\rm
MSPE}=\widehat{\rm MSPE}_{1}$ if something happens, and $\widehat{\rm
MSPE}=\widehat{\rm MSPE}_{2}$ otherwise. This strategy takes care of
the positivity issue, but it does not necessarily preserve the
second-order unbiasedness, even if $\widehat{\rm MSPE}_{1}$ and
$\widehat{\rm MSPE}_{2}$ are both second-order unbiased. In fact,
no rigorous proof has even been given that such a combined MSPE
estimator is both positive and second-order unbiased. In contrast,
there is no requirement that log-MSPE needs to be positive.
Therefore, for log-MSPE, one can simply focus on the second-order
unbiasedness of its estimator. Question is: How to obtain such an
estimator?

In the context of MSPE estimation, a standard approach is Prasad-Rao
(P-R) linearization (Prasad and Rao 1990). However, the approach is
not feasible to handle our current problem, which is much more
complicated. More specifically, we are interested in estimating the
log-MSPE when the small area predictor is obtained after a
model-selection procedure. The existing literature on inference
after model selection has mainly focused on the case of independent
observations (e.g., Rao and Wu 2001, sec. 12 and the references
therein, Leeb 2009, Berk, Brown and Zhao 2010). In particular, the
potential impact of model selection on MSPE has never been
rigorously addressed in the SAE literature. Intuitively, there is
an additional uncertainty involved in the model-selection process,
that needs to be taken into account in the MSPE estimation. The
P-R linearization method requires differentiability of the underlying
operation. This usually holds for standard estimation and prediction
procedures, but not for model selection. For example, the information
criteria, such as AIC (Akaike 1973) and BIC (Schwarz 1978), or the
fence methods (see Jiang 2014 for a review), select models from a
discrete space of candidate models. Even the shrinkage methods
(e.g., Tibshirani 1996, Fan and Li 2001) involve continuous but
non-differentiable penalty functions, such as the $L^{1}$ norm. See
M\"{u}ller, Scealy and Welsh (2013) for a review. Even if it is
possible to develop a P-R type method, the derivation is tedious, and
the final analytic expression is likely to be complicated. More
importantly, errors often occur in the process of derivations as well
as computer programming based on the lengthy expressions.

In this paper, we develop a unified jackknife approach that is
assisted by Monte-Carlo simulations for the estimation of log-MSPE.
As will be seen, the approach is applicable not just to the
current problem of SAE after model selection, but to a much broader
class of problems to obtain nearly unbiased estimators of quantities
that can be obtained via Monte-Carlo simulation, if one knows the
parameters that are involved. The method is especially attractive
if the quantity of interest does not carry a constraint, such as
non-negativity. This will be the case for the log-MSPE.
Furthermore, the Monte-Carlo jackknife method, called {\it McJack},
is ``one-formula-for-all'', which means that one needs not to
re-derive the formula, as in P-R type methods, every time there is
a new problem.

In the context of resampling methods, a well-known method is
jackknife-after-bootstrap (JAB; Efron 1992). There are major
differences between JAB and McJack. First, the objectives are
different. The main purpose of JAB is to assess accuracy of
the usual bootstrap estimates; while the objective of McJack
is to estimate quantities of interest, such as measures of
uncertainty for estimates based on the original data. Secondly,
JAB works, for the most part, under the standard nonparametric
bootstrap setting, to achieve efficient computation so
that no additional bootstrap samples are needed; in other
words, the JAB estimates are obtained from the original
bootstrap samples. However, this is difficult to do under a
parametric bootstrap setting. For example, although Efron
(1992) has discussed JAB with parametric bootstrap using the
idea of importance sampling, the approach does not necessarily
lead to a real gain in computation if the major computational
burden is not due to sampling. On the other hand, standard
nonparametric bootstrap procedures do not apply to SAE problems,
in spite of some variations that have been developed. See, for
example, Pfeffermann (2013), for a review. Finally, McJack does
not have to be associated with bootstrap--any kind of Monte-Carlo
method can be used to assist the computation. For example, Jiang,
Lahiri and Wan (2002; hereafter, JLW) discussed an example in
which the Monte-Carlo method used to compute the MSPE is not
considered as bootstrapping.

The rest of the paper is organized as follows. We begin by offering
a critical review of JLW, which has had significant impact in SAE.
We point out some undesirable features of JLW, and make two important
observations that lead to McJack. The latter is described in Section
3 with a theoretical justification. Estimation of log-MSPE in SAE
after model model selection is illustrated using an example. In
Section 4, we carry out simulation studies on performance of McJack,
and compare it with alternative approaches. A real data application
is considered in Section 5. We offer some discussion in Section 6.
Proofs of the theorems are given in Section 7.
\section{A brief review of JLW, and important observations}
\hspace{4mm}
In the context of resampling methods for SAE, Jiang, Lahiri, and
Wan (2002; hereafter, JLW) proposed a jackknife method for
estimating the MSPE of empirical best predictor (EBP) when the
parameters of interest are estimated by M-estimators. Let $\xi$
denote a mixed effect, for example, a small area mean. Let
$\tilde{\xi}$ and $\hat{\xi}$ denote the best predictor (BP), defined
as conditional expectation of $\xi$ given the data, $y$, and EBP of
$\xi$, respectively. Then, one has the decomposition:
\begin{eqnarray}
{\rm MSPE}(\hat{\xi})&=&{\rm MSPE}(\tilde{\xi})+{\rm E}\{(\hat{\xi}
-\tilde{\xi})^{2}\},
\end{eqnarray}
where ${\rm MSPE}(\hat{\xi})$ is defined as ${\rm E}\{(\hat{\xi}-
\xi)^{2}\}$ and ${\rm MSPE}(\tilde{\xi})$ is defined similarly. The idea
of JLW is to jackknife the two terms on the right side of (2) separately.
For the first term, the authors assume that it is a function of $\psi$,
a vector of parameters, that is, ${\rm MSPE}(\tilde{\xi})=b(\psi)$,
which can be computed analytically. The parameter vector $\psi$ is then
estimated by an M-estimator, defined as the solution, $\hat{\psi}$, to a
system of equations of the
following form:
\begin{eqnarray}
\sum_{i=1}^{m}f_{i}(\psi,y_{i})+a(\psi)&=&0.
\end{eqnarray}
In (3), $y_{i}$ is the data vector from the $i$th cluster (e.g., small
area), and the clusters are assumed to be independent; $f_{i}(\cdot,
\cdot)$ is a vector-valued function that satisfies ${\rm E}\{f_{i}(
\psi,y_{i})\}=0, 1\leq i\leq m$, if $\psi$ is the true parameter
vector; and $a(\cdot)$ corresponds to a penalizer, which in some cases
is the zero vector. The delete-$j$ estimator, $\hat{\psi}_{-j}$, of
$\psi$ is defined as the solution to the following system of equations:
\begin{eqnarray}
\sum_{i\neq j}f_{i}(\psi,y_{i})+a_{-j}(\psi)&=&0,
\end{eqnarray}
where $a_{-j}(\cdot)$ has a similar interpretation. Given the
M-estimators, $b(\psi)$ is estimated by a plug-in estimator, minus a
jackknife bias correction, that is,
\begin{eqnarray}
b(\hat{\psi})-\frac{m-1}{m}\sum_{j=1}^{m}\{b(\hat{\psi}_{-j})-b(\hat
{\psi})\}.
\end{eqnarray}
As for the second term on the right side of (2), it is estimated by
a jackknife variance-type estimator that has the following expression:
\begin{eqnarray}
\frac{m-1}{m}\sum_{j=1}^{m}(\hat{\xi}_{-j}-\hat{\xi})^{2},
\end{eqnarray}
where $\hat{\xi}_{-j}$ is a delete-j version of $\hat{\xi}$, the EBP,
defined in a certain way, which is not important for the current
paper. JLW showed that, when the two terms, (5) and (6), are put
together, the combined jackknife estimator of the MSPE of EBP is
second-order unbiased. The work has had a significant impact in
SAE, especially in the literature of resampling methods in SAE (e.g.,
Hall and Maiti 2006, Lohr and Rao 2009, Pfeffermann 2013, Rao and
Molina 2015). On the other hand, we note the following undesirable
features of JLW:\\
(a) JLW requires analytical computation of $b(\psi)$. More
specifically, JLW assumes posterior linearity, under which $b(\psi)$
has an analytic expression.\\
(c) JLW does not incorporate errors from model selection. In
particular, the proof for the second-order unbiased property of
(6) fails if a model selection procedure is involved prior to
obtaining the EBP, such as in Datta {\it et al.} (2011).\\
(c) JLW does not ensure a strictly positive MSPE estimator, in spite
of its second-order unbiasedness. See our discussion in Section 1
(5th paragraph).

As far as this paper is concerned, what is most important is not the
full JLW theory, but rather an intermediate result. In obtaining their
theory, JLW showed, in particular, that (5) is a second-order
unbiased estimator of $b(\psi)$, if the penalizers $a, a_{-j}, 1\leq
j\leq m$ in (3) and (4) satisfy certain mild conditions. In particular,
those conditions are satisfied if the penalizers are zero (vectors),
in which case the M-estimating equations are unbiased. Having given
the proof of the result, we realize the following two facts, both are
critically important to the idea of the current paper.

(I) The fact that $b(\psi)$ is an MSPE is not used anywhere in the
proof. In other words, as long as $b(\cdot)$ is a sufficiently
smooth function, and $\psi$ is estimated by the M-estimators, the
second-order unbiased estimation of $b(\psi)$ by (5) holds. In
particular, $b(\psi)$ can be $\log({\rm MSPE})$, which is of primary
interest here.

(II) More importantly, $b(\psi)$ does not have to have an analytic
expression, as long as one knows how to compute it. An analytic
expression would be nice, but, in the new era, the computation is
typically done by a computer, perhaps, a high-powered one. In
particular, suppose that, given $\psi$, $b(\psi)$ can be approximated
by a Monte-Carlo method to an arbitrary degree of accuracy. Then, one
can write a computer program, based on the Monte-Carlo, to compute
$b(\cdot)$ as a function. Given this ``computer-powered'' function,
all one needs to do is to plug the M-estimators, $\hat{\psi},
\hat{\psi}_{-j}, 1\leq j\leq m$, into this function to obtain the
second-order unbiased estimator of $b(\psi)$.

The importance of the above observations is that they apply
to virtually any kind of situation, not just the EBP. In
particular, the predictor, $\hat{\xi}$, can be much more
complicated than the EBP, such as an EBP obtained following a
model-selection procedure. Also, the decomposition (2), the
posterior linearity assumption, and (6) are altogether not needed
to apply these observations. In the next section, we propose a new
method based on the two important observations that addresses all of
the undesirable features of JLW noted above. Other complicated
situations, to which our idea may apply, include (i) regression
inference after variable selection (e.g., Leeb 2009); (ii) mixed model
prediction with non-normal random effect distribution (e.g., Lahiri
and Rao 1995); and (iii) shrinkage estimation/selection with data-driven
choice of regularization parameter (e.g., Pang, Lin and Jiang 2015).
\section{Monte-Carlo jackknife}
\hspace{4mm}
We first illustrate the method using an example of EBLUP under a
Fay-Herriot model, where the BIC (Schwarz 1978) is used to select the
fixed covariates as well as whether to include the area-specific random
effects. The model can be expressed in a way more convenient for the
model selection problem:
\begin{eqnarray}
y_{i}&=&x_{i}'\beta+\sqrt{A}\xi_{i}+e_{i},
\end{eqnarray}
$i=1,\dots,m$, where the components of $x_{i}$ are to be selected from
a set of candidate covariates; $\xi_{i}\sim N(0,1)$; if $A>0$, the random
effects are included in the model; if $A=0$, the random effects are
excluded from the model; $e_{i}\sim N(0,D_{i})$, where $D_{i}, 1\leq
i\leq m$ are known; and the $\xi_{i}$'s and $e_{i}$'s are independent.
Note that there have been further considerations regarding the choice
of the random effects; see, for example, Datta {\it et al.} (2011),
but here we focus on a simpler situation. Let $M_{\rm f}$ denote a
full model, under which $x_{i}$ is the vector that includes all of
the candidate covariates, and $A\geq 0$. Denote the $x_{i}$ under
$M_{\rm f}$ by $x_{{\rm f},i}$, and the corresponding $\beta$ by
$\beta_{\rm f}$. Let $\psi=(\beta_{\rm f}',A)'$. It is easy to see
that $M_{\rm f}$ is, at least, a correct model, which means that (7)
holds with $x_{i}$ replaced by $x_{{\rm f},i}$, $\beta$ replaced by
$\beta_{\rm f}$, and the range of $A$ being $[0,\infty)$. Of course,
some of the components of $\beta_{\rm f}$ may be zero, in case that
the full model can be simplified, and the true $A$ may be zero--these
are the reasons for the model selection. But this does not change the
fact $M_{\rm f}$ is a correct model. In particular, the true small-area
mean, $\theta_{i}$, can be expressed as
\begin{eqnarray}
\theta_{i}&=&x_{{\rm f},i}'\beta_{\rm f}+\sqrt{A}\xi_{i}.
\end{eqnarray}
On the other hand, under a candidate model, $M$, which corresponds to
(7), the EBLUP of $\theta_{i}$ can be expressed as
\begin{eqnarray}
\tilde{\theta}_{i}&=&\frac{\hat{A}}{\hat{A}+D_{i}}y_{i}+\frac
{D_{i}}{\hat{A}+D_{i}}x_{i}'\hat{\beta},
\end{eqnarray}
where $\hat{\beta}=\{\sum_{i=1}^{m}(\hat{A}+D_{i})^{-1}x_{i}x_{i}'
\}^{-1}\sum_{i=1}^{m}(\hat{A}+D_{i})^{-1}x_{i}y_{i}$, and $\hat{A}$
is a consistent estimator of $A$ obtained using a certain method (e.g.,
P-R, ML, REML; see Rao and Molina 2015). The BIC procedure chooses the
model, $M$, by minimizing
\begin{eqnarray}
{\rm BIC}(M)&=&-2\hat{l}+|M|\log(m),
\end{eqnarray}
where $\hat{l}$ is the maximized log-likelihood under $M$; $|M|={\rm
dim}(\beta)+1$ if $M$ includes the random effects, and $|M|={\rm 
dim}(\beta)$ if $M$ excludes the random effects. Here, for simplicity,
we assume that $X=(x_{i}')_{1\leq i\leq m}$ is full rank under any $M$.
Let the minimizer of (10) be $\hat{M}$. We then compute the EBLUP (9)
under $M=\hat{M}$, that is,
\begin{eqnarray}
\hat{\theta}_{i}&=&\frac{\hat{A}_{\hat{M}}}{\hat{A}_{\hat{M}}+
D_{i}}y_{i}+\frac{D_{i}}{\hat{A}_{\hat{M}}+D_{i}}x_{\hat{M},i}'\hat
{\beta}_{\hat{M}},
\end{eqnarray}
where $\hat{\beta}_{\hat{M}}$ and $\hat{A}_{\hat{M}}$ are the $\hat
{\beta}$ and $\hat{A}$ obtained under $\hat{M}$. The MSPE of
interest is
\begin{eqnarray}
{\rm MSPE}(\hat{\theta}_{i})&=&{\rm E}(\hat{\theta}_{i}-\theta_{i})^{2},
\end{eqnarray}
where $\theta_{i}$ is given by (8). It is clear that the
joint distribution of $(\theta_{i},y_{i}), 1\leq i\leq m$ depends
only on $\psi=(\beta_{\rm f}',A)$. Thus, (12) is a function of $\psi$
and so is its logarithm. Let
\begin{eqnarray}
b(\psi)&=&\log\{{\rm MSPE}(\hat{\theta}_{i})\}.
\end{eqnarray}
Given $\psi$, for the $k$th Monte-Carlo simulation, one first generates
$\theta_{i}$ by (8) with $\xi_{i}$ replaced by $\xi_{i}^{(k)}$, $1\leq
i\leq m$, generated independently from $N(0,1)$. Denote the generated
$\theta_{i}$ by $\theta_{i}^{(k)}$. Next, let $y_{i}^{(k)}=
\theta_{i}^{(k)}+e_{i}^{(k)}$, $1\leq i\leq m$, where $e_{i}^{(k)}\sim
N(0,D_{i}), 1\leq i\leq m$, generated independently and independent
with $\xi_{i}^{(k)}$'s. The Monte-Carlo approximation to $b(\psi)$ is
\begin{eqnarray}
\tilde{b}(\psi)&=&\log\left[\frac{1}{K}\sum_{k=1}^{K}\left\{\hat
{\theta}_{i}^{(k)}-\theta_{i}^{(k)}\right\}^{2}\right],
\end{eqnarray}
where $\hat{\theta}_{i}^{(k)}$ is obtained the same way as the $\hat
{\theta}_{i}$ of (11) except with $y_{i}$ replaced by $y_{i}^{(k)}$,
$1\leq i\leq m$. Write the above procedure as a function, say, $\tilde
{b}(\psi)={\bf mcjack}(\psi)$, that computes (14) for every given
$\psi$. Now suppse that $\hat{\psi}$ is an M-estimator of $\psi$. For
example, $\hat{A}$ is the P-R estimator (Prasad and Rao 1990; truncated
at zero if the expression turns out to be negative), and $\hat
{\beta}_{\rm f}$ is given below (9) with $x_{i}=x_{{\rm f},i}, 1\leq
i\leq m$. Let $\hat{\psi}_{-j}$ be the delete-$j$ version of $\hat
{\psi}$. The McJack estimator of (13) is then given by
\begin{eqnarray}
\widehat{b(\psi)}&=&\tilde{b}(\hat{\psi})-\frac{m-1}{m}\sum_{j=1}^{m}
\{\tilde{b}(\hat{\psi}_{-j})-\tilde{b}(\hat{\psi})\}.
\end{eqnarray}

Although the above illustration is based on the Fay-Herriot model,
its general principle, namely, (12)--(15), applies to much broader
cases. Using the result of JLW, we can justify the second-order
unbiasedness of McJack under the general framework. The justification
also takes into account effect of the Monte-Carlo errors. First note
that, to establish a rigorous result about the unbiasedness, we need
to make sure that the expectations of $\tilde{b}(\hat{\psi}_{-j}),
0\leq j\leq m$ exist. To avoid complicated technical conditions, we
regularize these estimators (e.g., Jiang {\it et al.} 2002, Das {\it
et al.} 2004). Let $\tilde{s}(\psi)=\exp\{\tilde{b}(\psi)\}$, and
define
\begin{eqnarray*}
\hat{s}(\psi)&=&\left\{\begin{array}{cc}e^{-\lambda m^{\rho}},&
{\rm if}\;\tilde{s}(\psi)<e^{-\lambda m^{\rho}},\\
\tilde{s}(\psi),&{\rm if}\;e^{-\lambda m^{\rho}}\leq \tilde{s}(\psi)
\leq e^{\lambda m^{\rho}},\\
e^{\lambda m^{\rho}},&{\rm if}\;\tilde{s}(\psi)>e^{\lambda m^{\rho}},
\end{array}\right.
\end{eqnarray*}
and $\hat{b}(\psi)=\log\{\hat{s}(\psi)\}$, where $\lambda,\rho$ are
given positive numbers. Let $s(\psi)$ denote ${\rm MSPE}(\hat
{\theta}_{i})$ when $\psi$ is the true parameter vector. We truncate
$s(\cdot)$ the same way as $\tilde{s}(\cdot)$, and let $b(\psi)=
\log\{s(\psi)\}$. For notation convenience, write $\hat{\psi}_{-0}=
\hat{\psi}$. Also, let $F_{-0}(\psi), F_{-j}(\psi)$ denote the left
sides of (3) and (4), respectively. The M-estimators, $\hat{\psi}_{-j},
0\leq j\leq m$ are said to be consistent uniformly (c.u.) at rate
$m^{-d}$ if, for any $\delta>0$, there is a constant $c_{\delta}$ such
that
\begin{eqnarray*}
{\rm P}(A_{j,\delta}^{c})&\leq&c_{\delta}m^{-d},\;\;\;0\leq j\leq m,
\end{eqnarray*}
where $A_{j,\delta}$ is the event that $F_{-j}(\hat{\psi}_{-j})=0$
and $|\hat{\psi}_{-j}-\psi|\leq\delta$, with $\psi$ being the true
parameter vector. Also, write $f_{i}=f_{i}(\psi,y_{i})$, $g_{i}=
\partial f_{i}/\partial\psi'$, $h_{i,k}=\partial^{2}f_{i,k}/
\partial\psi\partial\psi'$, where $f_{i,k}$ is the $k$th component
of $f_{i}$. Furthermore, for any function $f$ of $\psi$, define
$$\|\Delta^{3}f\|_{w}=\max_{1\leq s,t,u\leq r}\sup_{|\tilde{\psi}
-\psi|\leq w}\left|\frac{\partial^{3}f(\tilde{\psi})}{\partial
\psi_{s}\partial\psi_{t}\partial\psi_{u}}\right|,$$
where $\psi$ is the true parameter vector, and $r={\rm dim}(\psi)$.
A similar definition is extended to $\|\Delta^{4}f\|_{w}$. The
spectral norm of a matrix, $B$, is defined as $\|B\|=\sqrt{
\lambda_{\max}(B'B)}$, where $\lambda_{\max}$ denotes the largest
eigenvalue. Also write $\Delta_{j}=a-a_{-j}$, where $a, a_{-j}$ are
the functions of $\psi$ that appear in (3) and (4), respectively.
We shall consider estimation of log-MSPE of $\hat{\theta}_{i}$,
a predictor of $\theta_{i}$ after model selection, for a fixed $i$.
Furthermore, we assume that the
Monte-Carlo samples, under $\psi$, are generated by first
generating some standard [e.g., $N(0,1)$] random variables and
then plugging $\psi$. For example, under the full Fay-Herriot model
of (7), $y_{i}$ is generated by first generating the $\xi_{i}$'s
and $\eta_{i}$'s, which are independent $N(0,1)$, and then letting
$y_{i}=x_{{\rm f},i}'\beta_{\rm f}+\sqrt{A}\xi_{i}+\sqrt{D}_{i}
\eta_{i}$, with $\psi=(\beta_{\rm f}',A)'$. Let $\xi$ denote the
vector of the standard random variables. We first make the
following general assumptions.

{\it A1.} There are $d>2$ and $w>0$ such that the $2d$th moments of
$|f_{i}|$, $\|g_{i}\|$, $\|h_{i,k}\|$, $\|\Delta^{3}f_{i,k}\|_{w}$,
$1\leq i\leq m, 1\leq k\leq r$ are bounded for some $d>2+\rho$.

{\it A2.} For the same $d$ and $w$ in {\it A1}, $a_{-j}$ and its up
to third order partial derivatives, $0\leq j\leq m$, as well as
$\Delta_{j}, 1\leq j\leq m$, all evaluated at $\tilde{\psi}$, are
bounded uniformly for $|\tilde{\psi}-\psi|\leq w$, where $\psi$ is
the true parameter vector, and $m^{\tau}(|\Delta_{j}|\vee\|\partial
\Delta_{j}/\partial\psi\|), 1\leq j\leq m$, evaluated at $\psi$,
are bounded, where $\tau=(d-2)/(2d+1)$.

{\it A3.} The log-MSPE function $b(\cdot)$ of (13) is four-times
continuously differentiable, and, for the same $w$ in {\it A1},
$\|\Delta^{4}b\|_{w}$ is bounded.

{\it A4.} $\limsup_{m\rightarrow\infty}\|\{{\rm E}(\bar{g})\}^{-1}\|
<\infty$, where $\bar{g}=m^{-1}\sum_{j=1}^{m}g_{j}$, evaluated at
the true $\psi$.

{\it A5.} $\hat{\psi}_{-j}, 0\leq j\leq m$ are c.u. at rate $m^{-d}$
for the same $d$ in {\it A1}.

{\it A6.} $\sum_{j=1}^{m}\Delta_{j}=O(m^{-\nu})$ for some $\nu>0$.

Recall the way that the Monte-Carlo samples are generated specified
above {\it A1}. Under this assumption, $\theta_{i}^{(k)}, \hat
{\theta}_{i}^{(k)}, 1\leq k\leq K$, generated under $\tilde{\psi}$,
are functions of $\tilde{\psi}$ and $\xi$. The additional assumptions
below are regarding the Monte-Carlo sampling.

{\it A7.} $\xi$ is independent with the data, $y$.
 
{\it A8.} Let $\psi$ be the true parameter vector, and $w$ be the
same as in {\it A1}. There are constants $0<c_{1}<c_{2}$ such that
$c_{1}\leq s(\tilde{\psi})\leq c_{2}$ for $|\tilde{\psi}-\psi|\leq
w$, and random variables $G_{k}, 1\leq k\leq K$, which do not depend
on $\tilde{\psi}$, such that $|\hat{\theta}_{i}^{(k)}-
\theta_{i}^{(k)}|\leq G_{k}$ and ${\rm E}(G_{k}^{q})$ are bounded
for some $q\geq 2\{2+(\rho\vee 1)\}$.

{\it A9.} $m^{2}/K\rightarrow 0$, as $m\rightarrow\infty$.

{\bf Theorem 1.} Suppose that {\it A1}--{\it A9} hold. Let $\widehat
{b(\psi)}$ denote (15) with $\tilde{b}$ replaced by $\hat{b}$. Then,
we have ${\rm E}\{\widehat{b(\psi)}-b(\psi)\}=o(m^{-1})$, where $\psi$
is the true $\psi$ [hence $b(\psi)$ is the true log-MSPE], and ${\rm
E}$ is with respect to both $y$ and $\xi$.

The next result focuses on the special case of Fay-Herriot model.

{\bf Theorem 2.} Suppose that the true $A>0$, and there are positive
constants $0<c_{1}<c_{2}$ such that $c_{1}\leq|x_{{\rm f},i}|\leq c_{2}$,
$c_{1}\leq D_{i}\leq c_{2}, 1\leq i\leq m$. Furthermore, suppose that
\begin{eqnarray}
\limsup_{m\rightarrow\infty}\lambda_{\min}\left(\frac{1}{m}\sum_{i=1}^{m}
x_{{\rm f},i}x_{{\rm f},i}'\right)&>&0,
\end{eqnarray}
and {\rm A9} holds. Then, the conclusion of Theorem 1 holds.

The proofs of Theorem 1 and Theorem 2 are given in Section 7.
\section{Numerical demonstration and simulation study}
\subsection{A simple demonstration}
\hspace{4mm}
To begin with, let us consider a very simple situation, which may
be viewed as a special case of the Fay-Herriot model,
\begin{eqnarray}
y_{i}&=&x_{i}'\beta+v_{i}+e_{i},\;\;\;i=1,\dots,m,
\end{eqnarray}
where the components of $x_{i}$ consist of an intercept, a group
indicator, $x_{1,i}$, which is $0$ if $1\leq i\leq m_{1}=m/2$, and
$1$ if $m_{1}+1\leq i\leq m$, and potentially a third component,
$x_{2,i}$, which is generated from the $N(0,1)$ distribution, and
fixed throughout the simulation. There are two candidate models:
Model 1, which includes $x_{2,i}$, and Model 2: which does not
include $x_{2,i}$. The model selection is carried out by BIC
(Schwarz 1978).

For this demonstration, we consider a special case that the
variance of the random effects, $v_{i}$, is known to be zero, that
is, $A=0$. There have been considerations of such situations in SAE
(e.g., Datta {\it et al.} 2011). The variance of $e_{i}$, $D_{i}$,
is equal to $1$ for $1\leq i\leq m_{1}$, and $a$ for $m_{1}+1\leq
i\leq m$, where the value of $a$ is either $4$ or $16$. Because
$A=0$, the small area mean, $\theta_{i}$, under a given model, is
equal to $x_{i}'\beta$. The corresponding EBLUP is $\hat{\theta}_{i}
=x_{i}'\hat{\beta}$, where $\hat{\beta}=(X'D^{-1}X)^{-1}X'D^{-1}y$,
with $X=(x_{i}')_{1\leq i\leq m}$ and $D={\rm diag}(D_{i}, 1\leq
i\leq m)$, is the best linear unbiased estimator (BLUE) of $\beta$
(e.g., Jiang 2007, sec. 2.3), under the given model. Due to the
unbiasedness of the BLUE, the MSPE of the EBLUP is equal to its
variance, that is,
\begin{eqnarray}
{\rm MSPE}(\hat{\theta}_{i})={\rm var}(\hat{\theta}_{i})=x_{i}'(X'D^{-1}
X)^{-1}x_{i},\;\;\;1\leq i\leq m,
\end{eqnarray}
which are known under the given model. Now suppose that the EBLUP is
obtained based on the model selected by the BIC. A naive estimator of
the MSPE of $\hat{\theta}_{i}$, which ignores model selection, would
be (18) computed under the selected model. The naive estimator of
the log-MSPE is the logarithm of the naive MSPE estimator. We
compare this estimator with two competitors. The first is what we
call bootstrap MSPE estimator, which corresponds to the first term
in (15), that is, without the jackknife bias correction, where
$b(\cdot)$ is the log-MSPE function. The second is the McJack
estimator given by (15). The bootstrap and McJack estimators are
computed based on $K=1000$ Monte-Carlo samples.

A series of simulation studies were carried out with $m=20$ and
$\beta_{0}=\beta_{1}=1$, where $\beta_{0}$ is the intercept and
$\beta_{1}$ the slope of $x_{1,i}$, and under two different
true underlying models. In the first scenario, Model 1 is the true
underlying model with the slope of $x_{2,i}$, $\beta_{2}=0.5$. In
the second scenario, Model 2 is the true underlying model (i.e.,
$\beta_{2}=0$). We present the simulated percentage relative bias
(\%RB), based on $N_{\rm sim}=1000$ simulation runs, in Figures 2
and 3, where, for a given area, the \%RB is defined as
\begin{eqnarray}
\%{\rm RB}&=&\left[\frac{{\rm E}\{\widehat{\log({\rm MSPE})}\}
-\log({\rm MSPE})}{|\log({\rm MSPE})|}\right]\times 100\%,
\end{eqnarray}
MSPE is the true MSPE based on the simulations, and ${\rm E}\{\widehat
{\log({\rm MSPE})}\}$ is the mean of the estimated log-MSPE based on
the simulations.
\begin{figure}[pt]
\includegraphics[height=17.5cm,width=15cm]{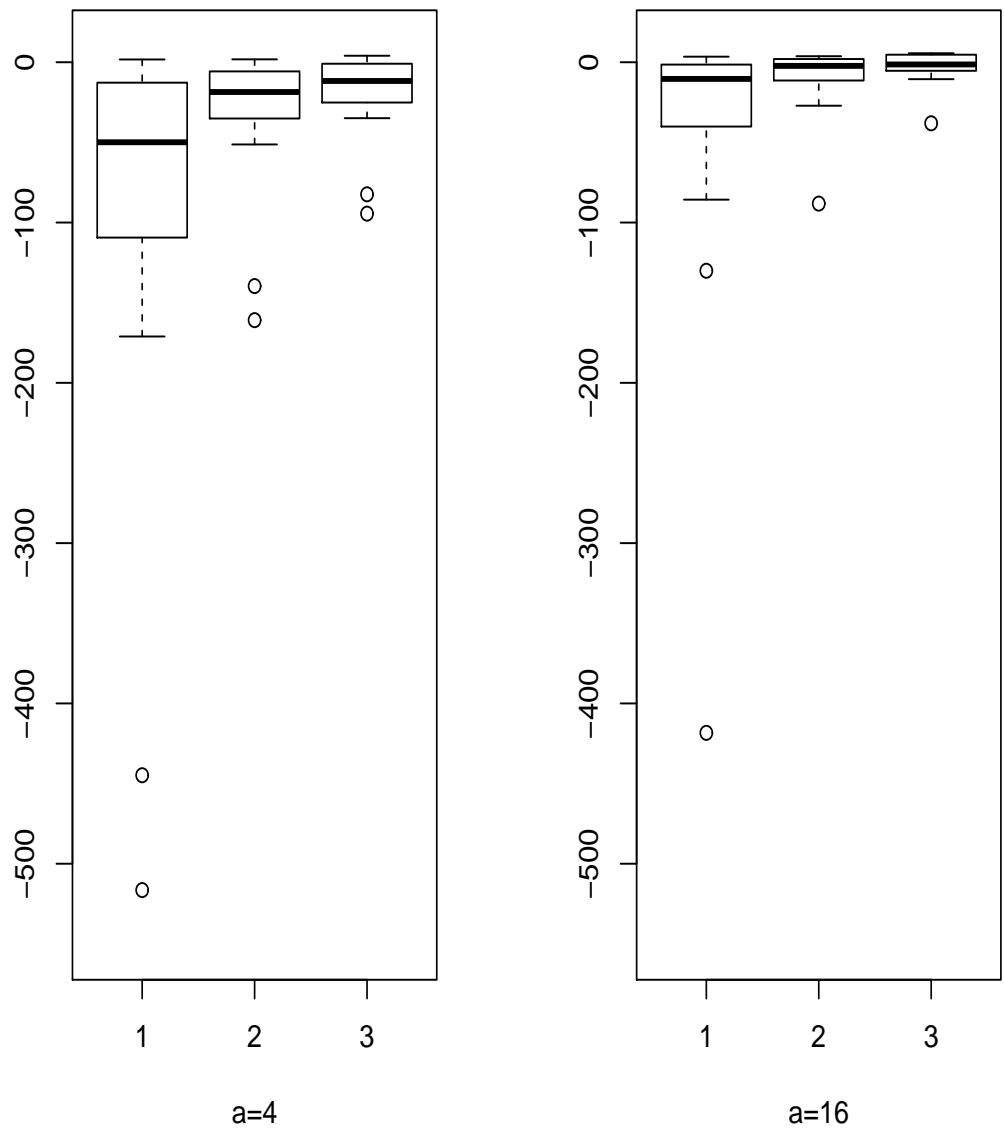}
\caption{\sl Boxplots of \%RB when Model 1 is the true model. In
each plot, from left to right: 1--Naive estimator, 2--bootstrap
estimator, and 3--McJack estimator, of log-MSPE.}
\end{figure}
\begin{figure}[pt]
\includegraphics[height=17.5cm,width=15cm]{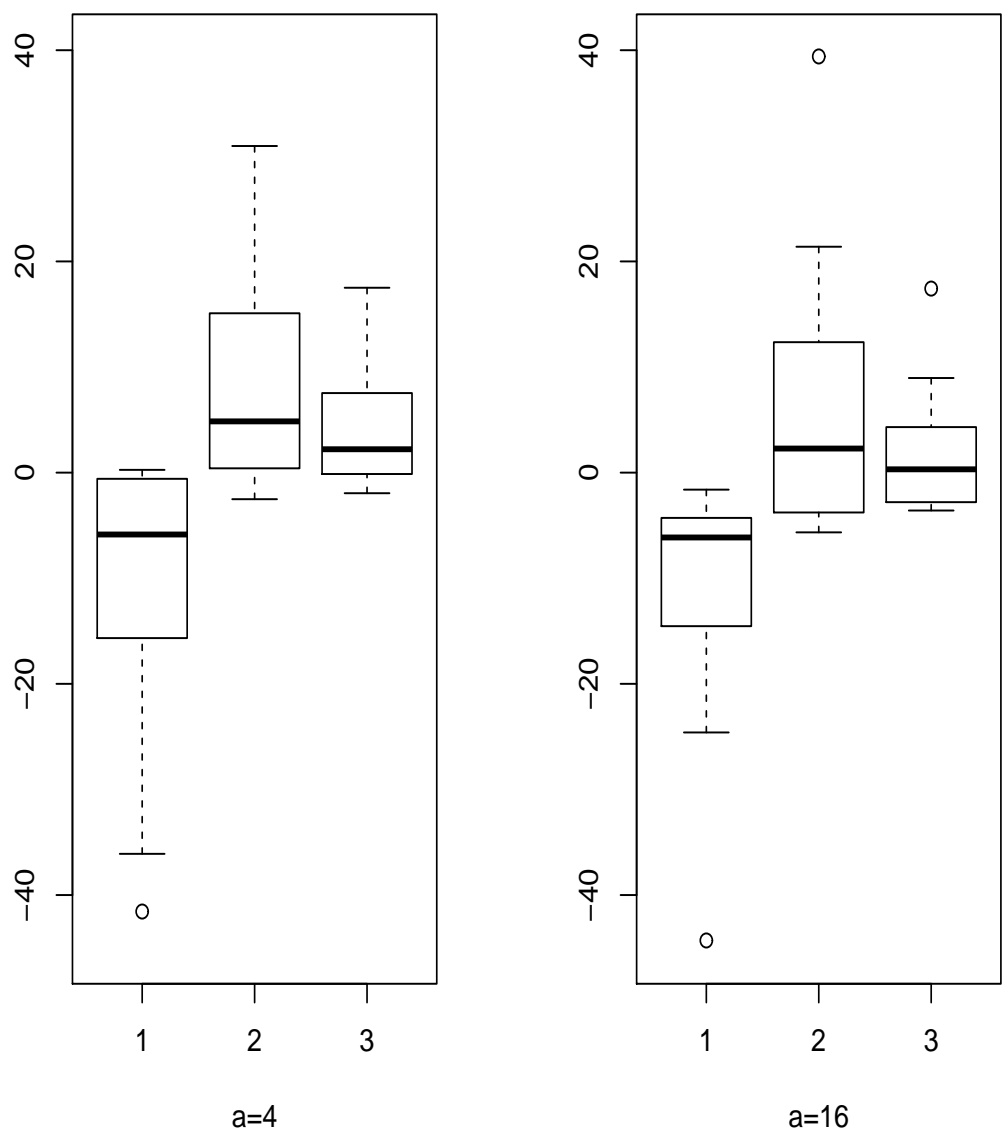}
\caption{\sl Boxplots of \%RB when Model 2 is the true model. In
each plot, from left to right: 1--Naive estimator, 2--bootstrap
estimator, and 3--McJack estimator, of log-MSPE.}
\end{figure}
It is seen that the naive estimator significantly under estimate the
log-MSPE; in fact, when Model 1 is the true model, the \%RB for one
of the areas is 516\% in the case of $a=4$, and there is a similar
case in the case of $a=16$. More specifically, there are some
interesting trend observed. Namely, when the true model is Model 1,
all of the methods seem to under-estimate the log-MSPE, but the
bootstrap and McJack estimators are doing much better, with McJack
offering significant improvement over the bootstrap. On the other hand,
when the true model is Model 2, the naive estimator again under-estimate
the log-MSPE, but the bootstrap and McJack estimators seem
to over-estimate the log-MSPE, with McJack significantly improving
the bootstrap. The amount of underestimation by the niave estimator
is less dramatic when Model 2 is the true model compared to when
Model 1 is the true model. One explanation is that the BIC is
known to have the tendency to over-penalize larger models. This
would have bigger impact when Model 1 is the true model, which is
the full model. In other words, there is a higher chance of model
misspecification by the BIC, which impacts the log-MSPE estimation.
To have a closer look at the numbers, we present one set of the
detailed results in Table 2.
\begin{table}
\begin{center}
\caption{\textbf{Log-MSPE estimation}: Model 2 is True Model; $a=4$;
\%RB in (\hspace{0.5mm})s.}
\begin{tabular}{c|cccc}
Area&True log-MSPE&E(Naive Est.)&E(Bootstrap Est.)&E(McJack Est.)\\ \hline
1&-1.98&-2.26 (-14.0)&-1.79 (9.6)&-1.91 (3.3)\\
2&-1.62&-2.21 (-36.1)&-1.22 (25.0)&-1.41 (12.8)\\
3&-2.07&-2.27 (-9.8)&-1.95 (5.6)&-2.01 (2.9)\\
4&-2.20&-2.30 (-4.3)&-2.26 (-2.5)&-2.25 (-1.9)\\
5&-1.70&-2.22 (-30.4)&-1.33 (21.7)&-1.52 (10.7)\\
6&-2.05&-2.27 (-10.8)&-1.91 (6.7)&-1.97 (3.7)\\
7&-2.14&-2.29 (-6.9)&-2.11 (1.5)&-2.16 (-1.0)\\
8&-1.55&-2.20 (-41.6)&-1.11 (28.4)&-1.28 (17.5)\\
9&-2.19&-2.30 (-4.8)&-2.23 (-1.6)&-2.22 (-1.4)\\
10&-2.06&-2.27 (-10.2)&-1.94 (6.0)&-2.00 (3.2)\\
11&-0.91&-0.92 (-0.5)&-0.91 (0.6)&-0.91 (-0.0)\\
12&-0.91&-0.92 (-0.1)&-0.91 (0.2)&-0.92 (-0.1)\\
13&-0.76&-0.89 (-17.4)&-0.61 (19.8)&-0.69 (9.5)\\
14&-0.87&-0.90 (-3.7)&-0.78 (10.4)&-0.82 (5.6)\\
15&-0.92&-0.92 (0.3)&-0.92 (0.1)&-0.92 (-0.1)\\
16&-0.92&-0.92 (0.1)&-0.92 (0.1)&-0.92 (-0.1)\\
17&-0.74&-0.88 (-18.1)&-0.52 (30.9)&-0.62 (17.2)\\
18&-0.92&-0.91 (0.1)&-0.90 (2.1)&-0.91 (1.1)\\
19&-0.91&-0.92 (-0.6)&-0.90 (0.7)&-0.91 (-0.0)\\
20&-0.88&-0.91 (-3.6)&-0.84 (4.1)&-0.87 (1.5)
\end{tabular}
\end{center}
\end{table}
\subsection{Testing the presence of random effects in a Fay-Herriot model}
\hspace{4mm}
Datta {\it et al.} (2011) proposed a method of model selection by
testing for the presence of the area-specific random effects, $v_{i}=
\sqrt{A}\xi_{i}$, in the Fay-Herriot model (7). This is equivalent to
testing the null hypothesis ${\rm H}_{0}: A=0$. The test statistic,
$T=\sum_{i=1}^{m}D_{i}^{-1}(y_{i}-x_{i}'\hat{\beta})^{2}$, where $\hat
{\beta}$ is the same as in Subsection 3.1, has a $\chi^{2}_{m-p}$
distribution, with $p={\rm rank}(X)$, under ${\rm H}_{0}$. If ${\rm H}_{0}$
is rejected, the EBLUP is used to estimate the small area mean $\theta_{i}$,
where in this simulation $A$ is estimated by the P-R estimator, and the
corresponding MSPE estimator is the P-R MSPE estimator; if ${\rm H}_{0}$ is
accepted, the estimator $\hat{\theta}_{i}=x_{i}'\hat{\beta}$ is used to
estimate $\theta_{i}$, and the corresponding MSPE is given by (18). Thus,
if the level of significance is chosen as $0.05$, the proposed MSPE
estimator, denoted by DHM, is the P-R MSPE estimator if
$T>\chi^{2}_{m-p}(0.05)$, and (18) if $T<=\chi^{2}_{m-p}(0.05)$.

We run a simulation study to compare the performance of McJack with DHM.
The simulation is under the full model considered in the previous
subsection (hence $p=3$), and three different true values of $A$: $A=0$,
$A=0.5$, and $A=1$. The boxplots of \%RB for these three cases are
presented in Figure 4, with the detailed numbers for DHM and McJack
given in Table 3. It is seen that DHM works better for the case $A=0$,
which is not surprising because, under the null hypothesis, the DHM
MSPE estimator is ``right'' 95\% of times. On the other hand, McJack
works significantly better in those two cases of nonzero $A$. Simple
simulations show that, in the latter cases, the probability of
rejecting the null hypothesis is about $0.26$ when $A=0.5$, and $0.44$
when $A=1$. The worst scenario seems to be the case where $A$ is not
zero but closer to zero ($A=0.5$). There are a few ``blown-up'' cases 
under this scenario where the \%RB exceeds 1000\% for DHM. It is also
obvious that McJack improves bootstrap in every case.
\begin{figure}[pt]
\includegraphics[height=17.5cm,width=15cm]{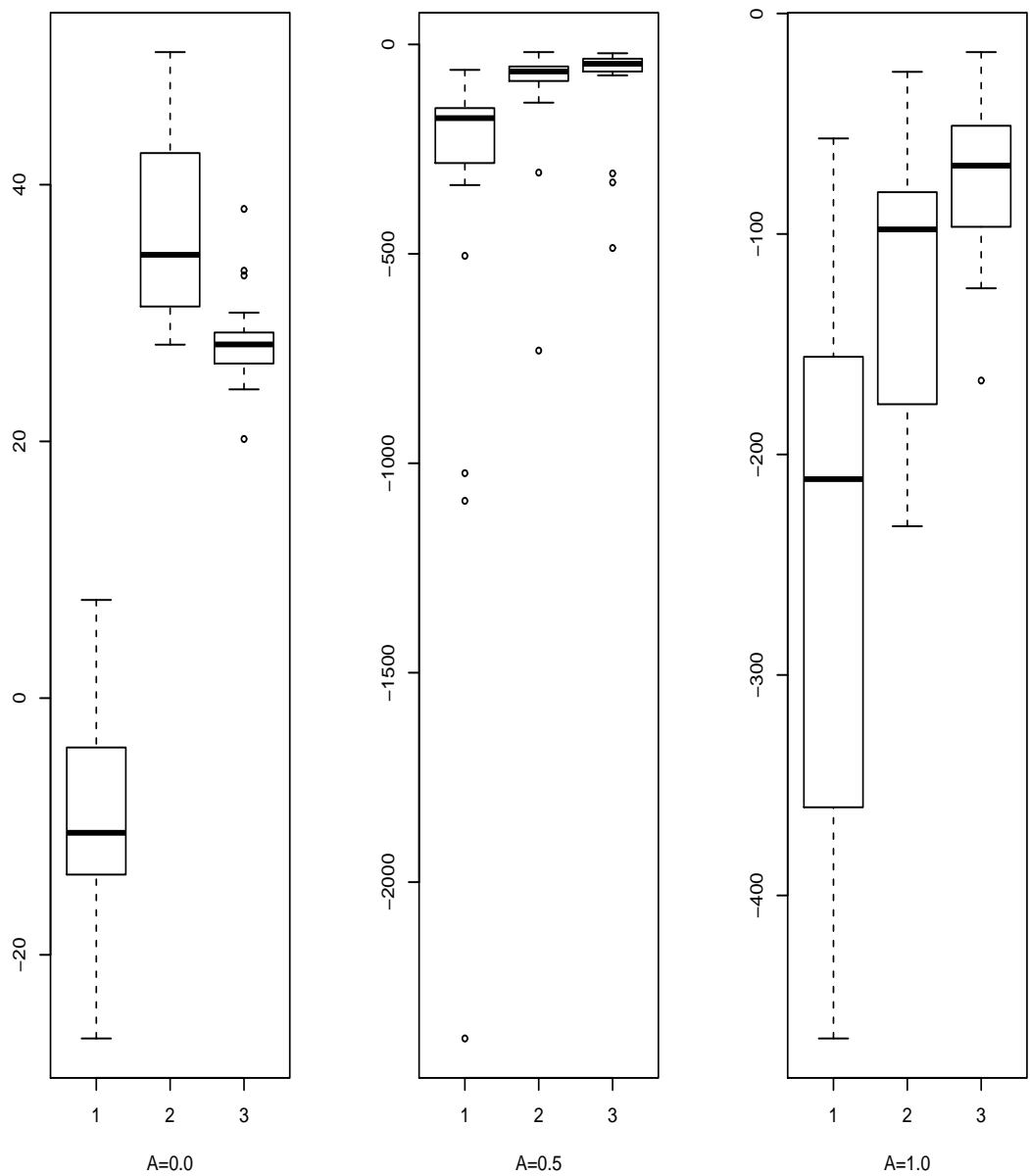}
\caption{\sl Boxplots of \%RB. In each plot, from left to right:
1--DHM, 2--bootstrap, 3--McJack. Scales are different due to the
huge difference in range.}
\end{figure}
\begin{table}
\begin{center}
\caption{\textbf{DHM vs McJack in \%RB}}
\begin{tabular}{c|rr|rr|rr}
Area&\multicolumn{2}{c|}{$A=0.0$}&\multicolumn{2}{c|}{$A=0.5$}&
\multicolumn{2}{c}{$A=1.0$}\\
&DHM&McJack&DHM&McJack&DHM&McJack\\ \hline
1&-13.6&26.2&-216.8&-59.8&-342.0&-99.5\\
2&0.3&26.0&-131.0&-45.8&-343.8&-72.9\\
3&-3.9&27.5&-107.4&-21.1&-135.8&-30.3\\
4&1.1&28.2&-158.0&-37.7&-362.9&-77.6\\
5&-8.1&24.9&-178.9&-36.9&-191.3&-59.6\\
6&-6.0&30.0&-180.3&-50.5&-375.4&-166.5\\
7&-3.1&24.1&-210.0&-51.5&-395.5&-124.6\\
8&7.6&27.6&-135.3&-33.1&-464.9&-123.0\\
9&-10.9&27.4&-149.4&-43.0&-357.2&-108.1\\
10&-3.8&28.6&-163.1&-31.4&-362.8&-94.0\\
11&-26.5&20.2&-60.8&-21.8&-220.3&-39.4\\
12&-10.7&33.3&-2373.3&-486.1&-210.4&-76.5\\
13&-13.9&27.5&-504.8&-74.0&-94.5&-18.4\\
14&-10.3&32.9&-173.1&-35.3&-188.2&-64.1\\
15&-17.5&25.7&-1023.6&-329.5&-163.0&-58.1\\
16&-4.4&38.1&-154.6&-46.6&-211.9&-65.2\\
17&-12.1&28.4&-335.8&-48.9&-197.6&-72.8\\
18&-11.4&27.1&-171.2&-22.0&-148.4&-59.9\\
19&-14.2&27.7&-230.6&-69.6&-56.7&-17.5\\
20&-18.4&28.3&-1089.6&-308.1&-104.1&-43.7
\end{tabular}
\end{center}
\end{table}

Another simulated example, in which the model selection is carried
out via a generalized information criterion (GIC) before the SAE,
is also considered. The details are deferred to Supplementary
Material due to the space limit.
\section{A real data example}
\hspace{4mm}
Morris and Christiansen (1995) presented a data set involving 23
hospitals (out of a total of 219 hospitals) that had at least 50
kidney transplants during a 27 month period (see Table 5). The
$y_{i}$'s are graft failure rates for kidney transplant operations, that
is, $y_{i}=$ number of graft failures$/n_{i}$, where $n_{i}$ is the number
of kidney transplants at hospital $i$ during the period of interest. The
variance for the graft failure rate, $D_{i}$, is approximated by
$(0.2)(0.8)/n_{i}$, where $0.2$ is the observed failure rate for all of
the hospitals. Thus, $D_{i}$ is assumed known. In addition, a severity
index, $s_{i}$, is available for each hospital, which is the average
fraction of females, blacks, children and extremely ill kidney
recipients at hospital $i$. Ganesh (2009) proposed a Fay-Herriot
model for the graft failure rates, which is (2) with $x_{i}'\beta=
\beta_{0}+\beta_{1}s_{i}$. Jiang {\it et al.} (2010) suggests that,
in a way, the optimal model for this data is a cubic model, that is,
(2) with $x_{i}'\beta=\beta_{0}+\beta_{1}s_{i}+\beta_{2}s_{i}^{2}+
\beta_{3}s_{i}^{3}$, which is also used in Datta {\it et al.} (2011).

We analyze the data under the latter model for the mean function but
with selection of the random effect factor using the strategy of Datta
{\it et al.} (2011), that is, by testing for the presence of the random
effects, $v_{i}$. At $\alpha=0.05$ level of significance, the test
statistic (see Subsection 3.2) $T=24.3$, while the critical value of
$\chi_{19}^{2}$ is $30.1$. Thus, the null hypothesis that $A=0$ is
not rejected. As a result, $\hat{\theta}_{i}=x_{i}'\hat{\beta}$ is
used as the estimate of $\theta_{i}$, according to Datta {\it et al.}
(2011). However, the main issue is how to assess the uncertainty. We
apply the three different methods investigated in Subsection 3.2 to
this data, and obtain the square roots of the estimated MSPEs, denoted
by DHM, BT, and MJ, respectively. Here the MSPE estimates are obtained
by taking the exponentials of the corresponding log-MSPE estimates.
The Monte-Carlo sample size for BT and MJ is $K=4000$. The results
are presented in Table 5. It is seen that the measures of uncertainty
by DHM are always smaller than those by BT and MJ. This is not
surprising because DHM does not take into account the potential
variation in model selection. As for the comparison between BT and
MJ, the latter measures are larger in most cases.
\begin{table}
\begin{center}
\caption{\textbf{The Hospital Data, Estimates, and Measures of
Uncertainty}}
{\scriptsize \setlength{\tabcolsep}{2.00mm}\small
\begin{tabular}{cccccccccc}\\
Area&$y_{i}$&$s_{i}$&$\sqrt{D_{i}}$&$\hat{\theta}_{i}$&DHM&BT&MJ&
$\tilde{\theta}_{i}$&MJ\\ \hline
1&.302&.112&.055&.221&.015&.029&.038&.238&.034\\
2&.140&.206&.053&.186&.013&.027&.019&.178&.019\\
3&.203&.104&.052&.214&.014&.029&.038&.215&.036\\
4&.333&.168&.052&.215&.011&.028&.044&.240&.040\\
5&.347&.337&.047&.349&.047&.047&.047&.349&.047\\
6&.216&.169&.046&.215&.011&.026&.030&.218&.024\\
7&.156&.211&.046&.183&.015&.027&.026&.176&.021\\
8&.143&.195&.046&.195&.011&.026&.032&.184&.034\\
9&.220&.221&.044&.177&.018&.029&.040&.186&.040\\
10&.205&.077&.044&.168&.015&.029&.048&.177&.049\\
11&.209&.195&.042&.195&.011&.026&.030&.199&.027\\
12&.266&.185&.041&.203&.010&.026&.029&.221&.026\\
13&.240&.202&.041&.189&.012&.026&.030&.203&.030\\
14&.262&.108&.036&.218&.014&.026&.021&.235&.018\\
15&.144&.204&.036&.188&.013&.025&.028&.174&.026\\
16&.116&.072&.035&.155&.017&.028&.038&.141&.042\\
17&.201&.142&.033&.228&.015&.025&.025&.221&.025\\
18&.212&.136&.032&.229&.015&.025&.025&.226&.025\\
19&.189&.172&.031&.213&.010&.023&.017&.205&.019\\
20&.212&.202&.029&.189&.012&.024&.038&.199&.034\\
21&.166&.087&.029&.189&.013&.024&.036&.180&.030\\
22&.173&.177&.027&.209&.010&.023&.032&.194&.034\\
23&.165&.072&.025&.155&.017&.022&.022&.159&.019\\
\end{tabular}}
\end{center}
\end{table}

As another comparison, we also computed the standard EBLUPs (i.e.,
without testing the presence of the random effects) and their
corresponding McJack estimates of $\sqrt{\rm MSPE}$. The results
are presented in the last two columns of Table 5, where $\tilde
{\theta}_{i}$ represents the EBLUPs and MJ the corresponding
estimated $\sqrt{\rm MSPE}$s. Note that the same data were also
analyzed by Datta {\it et al.} (2011), who stated that, because
the estimated MSPEs for DHM are much smaller than those for
EBLUP, the DHM method is ``significantly more accurate''. The
results of our analysis show that this is not necessarily the
case when additional variation in the model selection (by testing)
is taken into account, and estimated correctly: Out of the 23 small
areas, only 5 has smaller estimated $\sqrt{\rm MSPE}$ for DHM as
compared to EBLUP when comparing the MJs for both (column 8 vs
column 10).

Finally, there is one area, \#5, for which all of the uncertainty
measures give essentially the same results, 0.047 (although, to the
fourth digit, the DHM measure is still smaller than its BT and MJ
counterparts).
This case corresponds to the ``outlier'' for this data, according
to Jiang, Nguyen and Rao (2011). As noted by the latter authors
(also see Jiang {\it et al.} 2010), without this case, a quadratic, 
instead of cubic, mean function would fit the data well. However,
there is an over-fitting problem for this particular area, that is,
the outlier causes the cubic fit to be ``perfect'' for this area.
This means that the fitted cubic function goes through exactly the
data point; as a result, the direct estimate, $y_{5}$, is equal to
the regression estimate, $x_{5}'\hat{\beta}$. As a result, there is
no difference between the EBLUP and the direct and synthetic
estimates, regardless of the value of $D_{5}$ and how one estimates
$A$. Thus, in this case, every method essentially reduces to the
direct estimate, $y_{5}=0.347$, and its measure of uncertainty,
$\sqrt{D_{5}}=0.047$.

Another real-data example on estimation of median income of
four-person families is also considered. Again, the details are
deferred to Supplementary Material.
\section{Discussion}
\hspace{4mm}
We have shown that the impact of model selection in accuracy
measures may be complicated. If the accuracy measure only focuses
on the variance, model selection is likely to add additional
variation to the measure. This is shown, for example, in Subsection
3.1, where the EBLUP is an unbiased estimator, hence the MSPE
reduces to the variance. On the other hand, if the accuracy
measure is the MSPE, the overall impact of model selection
depends on the relative contributions of the bias and variance as
in the identity ${\rm MSPE}=({\rm prediction}\;{\rm bias})^{2}+{\rm
prediction}\;{\rm variance}$. As further discussed in Supplementary
Material, model selection helps to reduce the bias but this may be
at the cost of adding more variation. Because, in practice, it is
difficult to predict in which way, and how much, the overall impact
is, the best strategy is to obtain an accurate MSPE estimator. We
have shown that the latter can be done via McJack.
\section{Proofs}
\subsection{Proof of Theorem 1}
\hspace{4mm}
Throughout this proof, $\psi$ denotes the true parameter vector.
Let $\widetilde{b(\psi)}$ denote (15) with $\tilde{b}(\cdot)$
replaced by $b(\cdot)$. Also, $c$ denotes a positive, generic
constant, whose value may be different at different places. By
Theorem 5.2 of Jiang {\it et al.} (2002), we have
\begin{eqnarray}
{\rm E}_{y}\{\widetilde{b(\psi)}-b(\psi)\}&=&o(m^{-1-\gamma}),
\end{eqnarray}
where $\gamma=[(d-2)/(2d+1)]\wedge\nu>0$, and ${\rm E}_{y}$ denotes
expectation with respect to $y$. Because the left side of (20) does
not depend on $\xi$, the equation also holds with ${\rm E}_{y}$
replaced by ${\rm E}$.

Let ${\rm E}_{\xi}$ and ${\rm P}_{\xi}$ denote expectation and
probability with respect to $\xi$. Consider
\begin{eqnarray}
\widehat{b(\psi)}-\widetilde{b(\psi)}=\hat{b}(\hat{\psi})-
b(\hat{\psi})-\frac{m-1}{m}\sum_{j=1}^{m}\{\hat{b}(\hat{\psi}_{-j})
-b(\hat{\psi}_{-j})+b(\hat{\psi})-\hat{b}(\hat{\psi})\}.
\end{eqnarray}
Let $\tilde{\psi}$ be a fixed parameter vector such that $|\tilde
{\psi}-\psi|\leq w$. Then, we have
\begin{eqnarray}
\hat{b}(\tilde{\psi})-b(\tilde{\psi})&=&\{\hat{b}(\tilde{\psi})
-b(\tilde{\psi})\}1_{(c_{1}/2\leq\tilde{s}(\tilde{\psi})\leq 2c_{2})}+
\{\hat{b}(\tilde{\psi})-b(\tilde{\psi})\}1_{(\tilde{s}(\tilde{\psi})
<c_{1}/2)}\nonumber\\
&&+\{\hat{b}(\tilde{\psi})-b(\tilde{\psi})\}1_{(\tilde{s}(\tilde{\psi})
>2c_{2})}\nonumber\\
&=&I_{1}+I_{2}+I_{3}.
\end{eqnarray}
First note that, by {\it A8}, we have ${\rm P}_{\xi}\{\tilde{s}(\tilde
{\psi})<c_{1}/2\}\leq{\rm P}_{\xi}\{|\tilde{s}(\tilde{\psi})-s(\tilde
{\psi})|>c_{1}/2\}\leq(c_{1}/2)^{-q/2}{\rm E}_{\xi}\{|\tilde{s}(\tilde
{\psi})-s(\tilde{\psi})|^{q/2}\}$. Next, write $u_{k}=
\{\hat{\theta}_{i}^{(k)}-\theta_{i}^{(k)}\}^{2}$ and note that ${\rm
E}_{\xi}(u_{1})=s(\tilde{\psi})$. By Marcinkiewicz-Zygmund inequality
(e.g., Jiang 2010, p. 150), we have
\begin{eqnarray*}
{\rm E}_{\xi}\{|\tilde{s}(\tilde{\psi})-s(\tilde{\psi})|^{q/2}\}&=&
\frac{1}{K^{q/2}}{\rm E}_{\xi}\left[\left|\sum_{i=1}^{K}\{u_{k}-{\rm
E}_{\xi}(u_{1})\}\right|^{q/2}\right]\\
&\leq&\frac{c}{K^{q/2}}{\rm E}_{\xi}\left[\sum_{k=1}^{K}\{u_{k}-{\rm
E}_{\xi}(u_{1})\}^{2}\right]^{q/4}\\
&\leq&\frac{c}{K^{q/4}}\times\frac{1}{K}\sum_{k=1}^{K}{\rm E}_{\xi}[|
u_{k}-{\rm E}_{\xi}(u_{1})|^{q/2}]\\
&\leq&\frac{c}{K^{q/4}},
\end{eqnarray*}
using Jensen's inequality for the second-to-last step, and {\it A8}
for the last step. It follows, by {\it A8} and the definition of
$\hat{b}(\cdot), b(\cdot)$ that
\begin{eqnarray}
|{\rm E}_{\xi}(I_{2})|&\leq&cm^{\rho}K^{-q/4}.
\end{eqnarray}
By essentially the same argument, we also have
\begin{eqnarray}
|{\rm E}_{\xi}(I_{3})|&\leq&cm^{\rho}K^{-q/4}.
\end{eqnarray}

Now suppose that $c_{1}/2\leq\tilde{s}(\tilde{\psi})\leq 2c_{2}$.
We also know that $c_{1}\leq s(\tilde{\psi})\leq c_{2}$ by {\it A8}.
Thus, for sufficiently large $m$, we have $\hat{b}(\tilde{\psi})=
\tilde{b}(\tilde{\psi})$. By Taylor series expansion, we have
\begin{eqnarray}
\hat{b}(\tilde{\psi})-b(\tilde{\psi})&=&\tilde{b}(\tilde{\psi})-
b(\tilde{\psi})\nonumber\\
&=&\log\{\tilde{s}(\tilde{\psi})\}-\log\{s(\tilde{\psi})\}\nonumber\\
&=&\frac{\tilde{s}(\tilde{\psi})-s(\tilde{\psi})}{s(\tilde
{\psi})}-\frac{\{\tilde{s}(\tilde{\psi})-s(\tilde{\psi})\}^{2}}{2
s(\tilde{\psi})^{2}}+\frac{\{\tilde{s}(\tilde{\psi})-s(\tilde
{\psi})\}^{3}}{3\eta^{3}},
\end{eqnarray}
where $\eta$ lies between $s(\tilde{\psi})$ and $\tilde{s}(\tilde
{\psi})$; hence, we have $\eta\geq c_{1}/2$. It follows that
\begin{eqnarray}
\left|{\rm E}_{\xi}\left[\frac{\{\tilde{s}(\tilde{\psi})-s(\tilde
{\psi})\}^{3}}{3\eta^{3}}1_{(c_{1}/2\leq\tilde{s}(\tilde{\psi})\leq
2c_{2})}\right]\right|\leq\frac{8}{3c_{1}^{3}}{\rm E}_{\xi}\{|\tilde
{s}(\tilde{\psi})-s(\tilde{\psi})|^{3}\}\leq cK^{-3/2},
\end{eqnarray}
using an earlier inequality. Similarly, we have
\begin{eqnarray}
\left|{\rm E}_{\xi}\left[\frac{\{\tilde{s}(\tilde{\psi})-s(\tilde
{\psi})\}^{2}}{2s(\tilde{\psi})^{2}}1_{(c_{1}/2\leq\tilde{s}(\tilde
{\psi})\leq 2c_{2})}\right]\right|\leq cK^{-1}.
\end{eqnarray}
Furthermore, note that ${\rm E}_{\xi}\{\tilde{s}(\tilde{\psi})-s(\tilde
{\psi})\}=0$, thus, we have
\begin{eqnarray*}
\left|{\rm E}_{\xi}\left[\frac{\tilde{s}(\tilde{\psi})-s(\tilde
{\psi})}{s(\tilde{\psi})}1_{(c_{1}/2\leq\tilde{s}(\tilde{\psi})\leq
2c_{2})}\right]\right|&=&\frac{1}{s(\tilde{\psi})}\left|{\rm E}_{\xi}[
\{\tilde{s}(\tilde{\psi})-s(\tilde{\psi})\}1_{(\tilde{s}(\tilde{\psi})
<c_{1}/2\;{\rm or}\;\tilde{s}(\tilde{\psi})>2c_{2})}]\right|\\
&\leq&\frac{{\rm E}_{\xi}[\{\tilde{s}(\tilde{\psi})+s(\tilde{\psi})
\}1_{(\tilde{s}(\tilde{\psi})<c_{1}/2)}]}{s(\tilde{\psi})}\\
&&+\frac{{\rm E}_{\xi}[\{\tilde{s}(\tilde{\psi})+s(\tilde{\psi})
\}1_{(\tilde{s}(\tilde{\psi})>2c_{2})}]}{s(\tilde{\psi})}.
\end{eqnarray*}
By H\"{o}lder and Jensen's inequalities, {\it A8} and an earlier
result, we have
\begin{eqnarray*}
&&{\rm E}_{\xi}[\{\tilde{s}(\tilde{\psi})+s(\tilde{\psi})\}1_{(\tilde
{s}(\tilde{\psi})<c_{1}/2)}]\\
&\leq&[{\rm E}_{\xi}\{\tilde{s}(\tilde{\psi})+s(\tilde{\psi})
\}^{q/2}]^{2/q}[{\rm P}_{\xi}\{\tilde{s}(\tilde{\psi})<c_{1}/2
\}]^{1-2/q}\\
&\leq&c\left\{\frac{1}{K}\sum_{k=1}^{K}{\rm E}_{\xi}(u_{k}^{q/2})
+c_{2}^{q/2}\right\}^{2/q}K^{-(q/4)(1-2/q)}\\
&\leq&cK^{-(q-2)/4}.
\end{eqnarray*}
Similarly, we have ${\rm E}_{\xi}[\{\tilde{s}(\tilde{\psi})+s(\tilde
{\psi})\}1_{(\tilde{s}(\tilde{\psi})>2c_{2})}]\leq cK^{-(q-2)/4}$. It
follows that
\begin{eqnarray}
\left|{\rm E}_{\xi}\left[\frac{\tilde{s}(\tilde{\psi})-s(\tilde
{\psi})}{s(\tilde{\psi})}1_{(c_{1}/2\leq\tilde{s}(\tilde{\psi})\leq
2c_{2})}\right]\right|\leq cK^{-(q-2)/4}.
\end{eqnarray}
Combining (24)--(28), and the fact that $(q-2)/4\geq 1$ by {\it A8},
we conclude that
\begin{eqnarray}
|{\rm E}_{\xi}(I_{1})|&\leq&cK^{-1}.
\end{eqnarray}
Thus, combining (22)--(24), and (29), we have
\begin{eqnarray}
|{\rm E}_{\xi}\{\hat{b}(\tilde{\psi})-b(\tilde{\psi})\}|\leq
c\left[m^{\rho}K^{-q/4}+K^{-1}\right],\;\;{\rm if}\;|\tilde{\psi}-
\psi|\leq w,
\end{eqnarray}
where $c$ does not depend on $\tilde{\psi}$.

Now, for any $0\leq j\leq m$, we have
$${\rm E}\{\hat{b}(\hat{\psi}_{-j})-b(\hat{\psi}_{-j})\}={\rm
E}_{y}[{\rm E}_{\xi}\{\hat{b}(\hat{\psi}_{-j})-b(\hat{\psi}_{-j})|
\hat{\psi}_{-j}\}]={\rm E}_{y}\{\Delta(\hat{\psi}_{-j})\},$$
where $\Delta(\tilde{\psi})={\rm E}_{\xi}\{\hat{b}(\hat{\psi}_{-j})
-b(\hat{\psi}_{-j})|\hat{\psi}_{-j}=\tilde{\psi}\}={\rm E}_{\xi}\{
\hat{b}(\tilde{\psi})-b(\tilde{\psi})\}$ by {\it A7}. Thus, we have
\begin{eqnarray}
{\rm E}\{\hat{b}(\hat{\psi}_{-j})-b(\hat{\psi}_{-j})\}={\rm E}_{y}\{
\Delta(\hat{\psi}_{-j})1_{(|\hat{\psi}_{-j}-\psi|\leq w)}\}+{\rm E}_{y}\{
\Delta(\hat{\psi}_{-j})1_{(|\hat{\psi}_{-j}-\psi|>w)}\}.
\end{eqnarray}
By (30), the first term on the right side of (31) is bounded in
absolute value by $c[m^{\rho}K^{-q/4}+K^{-1}]$. As for the second
term, by the definition of $\hat{b}(\cdot), b(\cdot)$, and {\it A5},
it is bounded in absolute value by $cm^{\rho-d}$. Thus, in conclusion,
we have
\begin{eqnarray}
|{\rm E}\{\hat{b}(\hat{\psi}_{-j})-b(\hat{\psi}_{-j})\}|\leq c\left[
m^{\rho}K^{-q/4}+K^{-1}+m^{\rho-d}\right],\;\;0\leq j\leq m.
\end{eqnarray}

Combining (21), (32), we have
\begin{eqnarray}
|{\rm E}\{\widehat{b(\psi)}-\widetilde{b(\psi)}\}|\leq c\left(
m^{1+\rho}K^{-q/4}+\frac{m}{K}+m^{1+\rho-d}\right)=o(m^{-1}),
\end{eqnarray}
by {\it A9} and the conditions on $d, q$.

The result then follows by (20) (with ${\rm E}_{y}$ replaced by E)
and (33).
\subsection{Proof of Theorem 2}
\hspace{4mm}
First, by (i)--(iv) of Jiang {\it et al.} (2002, p. 1803), it is
easy to see that assumptions {\it A1}--{\it A6} are satisfied.
Assumption {\it A7} is satisfied by the statement above {\it A1}.
Thus, all we need is to verify assumption {\it A8}. Once again, in the
arguments below, $c$ denotes a positive constant whose value may be
different at different places.

Suppose that the data are generated under the parameter vector $\tilde
{\psi}$. Let $\tilde{\theta}_{i}$ denote the BP of $\theta_{i}$. Then,
we have $s(\tilde{\psi})={\rm MSPE}_{\tilde{\psi}}(\hat{\theta}_{i})
={\rm MSPE}_{\tilde{\psi}}(\tilde{\theta}_{i})+{\rm E}_{\tilde
{\psi}}\{(\hat{\theta}_{i}-\tilde{\theta}_{i})^{2}\}
\geq{\rm MSPE}_{\tilde{\psi}}(\tilde{\theta}_{i})
=\tilde{A}D_{i}/(\tilde{A}+D_{i})$.
Thus, if $0<A/2\leq\tilde{A}\leq 2A$, where $\tilde{A}$ is the $A$
component of $\tilde{\psi}$, and $A$ is the true $A$, $s(\tilde{\psi})$
is clearly bounded away from zero.

On the other hand, we have $s(\tilde{\psi})={\rm E}_{\tilde{\psi}}(
\hat{\theta}_{i}^{2})-2{\rm E}_{\tilde{\psi}}(\hat{\theta}_{i}
\theta_{i})+{\rm E}_{\tilde{\psi}}(\theta_{i}^{2})$. By (8), we
have ${\rm E}(\theta_{i}^{2})\leq 2(|x_{{\rm f},i}|^{2}|\tilde
{\beta}_{\rm f}|^{2}+\tilde{A}^{2})\leq c$, if, say $|\tilde{\beta}_{\rm
f}-\beta_{\rm f}|\leq 1$ and $\tilde{A}\leq 2A$. Also, by (11), and
Jensen's inequality, we have
\begin{eqnarray}
\hat{\theta}_{i}^{2}\leq\frac{\hat{A}_{\rm f}}{\hat{A}_{\rm f}+
D_{i}}y_{i}^{2}+\frac{D_{i}}{\hat{A}_{\rm f}+D_{i}}|x_{{\rm f},
i}\hat{\beta}_{\rm f}|^{2}\leq y_{i}^{2}+|x_{{\rm f},i}|^{2}|\hat
{\beta}_{\rm f}|^{2},
\end{eqnarray}
and, by (7), ${\rm E}_{\tilde{\psi}}(y_{i}^{2})=\{{\rm E}_{\tilde
{\psi}}(y_{i})\}^{2}+{\rm var}_{\tilde{\psi}}(y_{i})=(x_{{\rm f},i}
\tilde{\beta}_{\rm f})^{2}+\tilde{A}+D_{i}\leq|x_{{\rm f},i}|^{2}
|\tilde{\beta}_{\rm f}|^{2}+\tilde{A}+D_{i}\leq c$. Define
$P_{\rm f}=I_{m}-D^{-1/2}X_{\rm f}(X_{\rm f}'D^{-1}X_{\rm f})^{-1}
X_{\rm f}'D^{-1/2}$.
By Lemma 1 of Jiang (2000), with $\hat{V}=\hat{A}I_{m}+D$,
$D={\rm diag}(D_{i}, 1\leq i\leq m)$, $X_{D}=D^{-1/2}X_{\rm f}$,
$Z=D^{-1/2}$, $\Gamma=\hat{A}I_{m}$, and $\zeta=y-X_{\rm f}\tilde
{\beta}_{\rm f}$, we have
\begin{eqnarray}
\hat{\beta}_{\rm f}&=&(X_{\rm f}'\hat{V}^{-1}X_{\rm f})^{-1}X_{\rm
f}'\hat{V}^{-1}y\nonumber\\
&=&\tilde{\beta}_{\rm f}+(X_{\rm f}'\hat{V}^{-1}X_{\rm f})^{-1}X_{\rm
f}'\hat{V}^{-1}\zeta\nonumber\\
&=&\tilde{\beta}_{\rm f}
+\{X_{D}'(I_{m}+Z\Gamma Z')^{-1}X_{D}\}^{-1}X_{D}'(I_{m}+Z\Gamma
Z')^{-1}D^{-1/2}\zeta\nonumber\\
&=&\tilde{\beta}_{\rm f}
+(X_{D}'X_{D})^{-1}X_{D}'\{I_{m}-\hat{A}D^{-1}P_{\rm f}(I_{m}+\hat
{A}P_{\rm f}D^{-1}P_{\rm f})^{-1}\}D^{-1/2}\zeta\nonumber\\
&=&\tilde{\beta}_{\rm f}+(X_{\rm f}'D^{-1}X_{\rm f})^{-1}X_{\rm
f}'D^{-1}\zeta\nonumber\\
&&-(X_{\rm f}'D^{-1}X_{\rm f})^{-1}X_{\rm f}'D^{-1}\hat{A}D^{-1/2}P_{\rm
f}(I_{m}+\hat{A}P_{\rm f}D^{-1}P_{\rm f})^{-1}D^{-1/2}\zeta\nonumber\\
&=&\tilde{\beta}_{\rm f}+I_{1}-I_{2}.
\end{eqnarray}
Note that ${\rm E}_{\tilde{\psi}}(\zeta\zeta')=\tilde{A}I_{m}+D\leq
cD$ under the assumptions. Thus, we have
\begin{eqnarray}
{\rm E}_{\tilde{\psi}}(|I_{1}|^{2})&=&{\rm E}_{\tilde{\psi}}\left[{\rm
tr}\{(X_{\rm f}'D^{-1}X_{\rm f})^{-1}X_{\rm f}'D^{-1}\zeta\zeta'D^{-1}
X_{\rm f}(X_{\rm f}'D^{-1}X_{\rm f})^{-1}\}\right]\nonumber\\
&=&{\rm tr}\left\{(X_{\rm f}'D^{-1}X_{\rm f})^{-1}X_{\rm f}'D^{-1}{\rm
E}_{\tilde{\psi}}(\zeta\zeta')D^{-1}X_{\rm f}(X_{\rm f}'D^{-1}X_{\rm
f})^{-1}\right\}\nonumber\\
&\leq&c{\rm tr}\{(X_{\rm f}'D^{-1}X_{\rm f})^{-1}\}\nonumber\\
&\leq&\frac{c}{m\lambda_{\min}(m^{-1}X_{\rm f}'X_{\rm f})}.
\end{eqnarray}
Furthermore, we have
$$|I_{2}|\leq\|(X_{\rm f}'D^{-1}X_{\rm f})^{-1}X_{\rm f}'D^{-1}\|\cdot
\|\hat{A}D^{-1/2}P_{\rm f}(I_{m}+\hat{A}P_{\rm f}D^{-1}P_{\rm f})^{-1}
\|\cdot|D^{-1/2}\zeta|.$$
By a similar argument as above, we have
$$\|(X_{\rm f}'D^{-1}X_{\rm f})^{-1}X_{\rm f}'D^{-1}\|^{2}\leq\frac
{c}{m\lambda_{\min}(m^{-1}X_{\rm f}'X_{\rm f})}.$$
Next, let $\lambda_{1}\geq\dots\geq\lambda_{m}\geq 0$ be the eigenvalues
$P_{\rm f}D^{-1}P_{\rm f}$. Then, we have
$$\|\hat{A}D^{-1/2}P_{\rm f}(I_{m}+\hat{A}P_{\rm f}D^{-1}P_{\rm f})^{-1}
\|^{2}=\max_{1\leq i\leq m}\frac{\hat{A}^{2}\lambda_{i}}{(1
+\hat{A}\lambda_{i})^{2}}.$$
If $\hat{A}\lambda_{i}=0$, then $\hat{A}^{2}\lambda_{i}/(1+\hat
{A}\lambda_{i})^{2}=0$; otherwise, $\hat{A}^{2}\lambda_{i}/(1
+\hat{A}\lambda_{i})^{2}\leq\hat{A}^{2}\lambda_{i}/\hat{A}^{2}
\lambda_{i}^{2}=1/\lambda_{i}$. It follows that
$$\max_{1\leq i\leq m}\frac{\hat{A}^{2}\lambda_{i}}{(1+\hat
{A}\lambda_{i})^{2}}\leq\frac{1}{\lambda_{r}},$$
where $r={\rm rank}(P_{\rm f}D^{-1}P_{\rm f})$. Because $P_{\rm
f}D^{-1}P_{\rm f}u=0$ if and only if $P_{\rm f}u=0$, we have
$r={\rm rank}(P_{\rm f})$, and $P_{\rm f}$ is a projection matrix,
whose eigenvalues are $0$ or $1$. Also, because
$$P_{\rm f}D^{-1}P_{\rm f}\geq\frac{P_{\rm f}^{2}}{\max_{1\leq i\leq
m}D_{i}}=\frac{P_{\rm f}}{\max_{1\leq i\leq m}D_{i}},$$
by a well-known eigenvalue inequality (e.g., DasGupta 2008, p. 669),
we have
$$\lambda_{r}\geq\lambda_{r}\left(\frac{P_{\rm f}}{\max_{1\leq i\leq
m}D_{i}}\right)=\frac{\lambda_{r}(P_{\rm f})}{\max_{1\leq i\leq m}D_{i}}
=\frac{1}{\max_{1\leq i\leq m}D_{i}}.$$
Thus, in conclusion, we have
$$\|\hat{A}D^{-1/2}P_{\rm f}(I_{m}+\hat{A}P_{\rm f}D^{-1}P_{\rm
f})^{-1}\|^{2}\leq\max_{1\leq i\leq m}D_{i}.$$
Finally, it is easy to show that ${\rm E}_{\tilde{\psi}}(|D^{-1/2}
\zeta|^{2})\leq cm$. Thus, combining the results, we have
\begin{eqnarray}
{\rm E}_{\tilde{\psi}}(|I_{2}|^{2})&\leq&\frac{c}{\lambda_{\min}
(m^{-1}X_{\rm f}'X_{\rm f})}.
\end{eqnarray}
The upper bound for $s(\tilde{\psi})$ follows from (34)--(37).

The last part of {\it A8} follows from the above arguments by noting
that $\theta_{i}^{(k)}=x_{{\rm f},i}'\tilde{\beta}_{\rm f}+\tilde
{A}^{1/2}\xi_{i}^{(k)}$, $y_{i}=\theta_{i}^{(k)}+\sqrt{D_{i}}
\eta_{i}^{(k)}$, and $\xi_{i}^{(k)}, \eta_{i}^{(k)}$ are $N(0,1)$
random variables.

\vspace{2mm}

{\bf Acknowledgements.} The research of Jiming Jiang is partially
supported by the NSF grant SES-1121794. The research of Thuan
Nguyen is partially supported by the NSF grant SES-1118469. The
research of Jiming Jiang and Thuan Nguyen are partially supported by
the NIH grant R01-GM085205A1.

\end{document}